\documentclass[aps,prc,floatfix,nofootinbib,showpacs,superscriptaddress,twocolumn]{revtex4}

\usepackage[utf8]{inputenc}
\usepackage{gensymb}
\usepackage{latexsym}
\usepackage{latexsym}
\usepackage{amssymb}
\usepackage{amsmath}
\usepackage{graphicx}
\usepackage{amsfonts}
\usepackage[mathscr,scaled=1.15]{urwchancal}
\DeclareFontFamily{OT1}{pzc}{}
\DeclareFontShape{OT1}{pzc}{m}{it}%
{<-> s * [1.15] pzcmi7t}{}
\DeclareMathAlphabet{\mathpzc}{OT1}{pzc}{m}{it}
\usepackage{supertabular}
\usepackage{placeins}
\usepackage{graphicx}
\usepackage{graphics}
\usepackage{bm}
\usepackage{color}
\usepackage{graphics}
\usepackage{epsfig}
\usepackage{float}
\usepackage{subfigure}
\usepackage{soul}
\definecolor{purple}{rgb}{0.5,0,0.5}
\definecolor{blue}{rgb}{0.0,0,0.9}
\usepackage[colorlinks=true, pdfstartview=FitV, linkcolor=purple, citecolor= purple, urlcolor=blue]{hyperref}

\begin{document}
	
	\title{Contribution of neutral pseudoscalar mesons to $a_\mu^{\textrm{HLbL}}$ within a Schwinger-Dyson 
		equations approach to QCD}
	
	\author{Kh\'epani Raya}
	\email{khepani@nankai.edu.cn}
	\affiliation{School of Physics, Nankai University, Tianjin 300071, China}
	\author{Adnan Bashir}
	\email{adnan.bashir@umich.mx}
	\affiliation{Instituto de F\'{i}sica y Matem\'aticas, Universidad
		Michoacana de San Nicol\'as de Hidalgo, Morelia, Michoac\'an
		58040, M\'{e}xico.}
	\author{Pablo Roig}
	\email{proig@fis.cinvestav.mx}
	\affiliation{Centro de Investigaci\'on y de Estudios Avanzados, Apartado \\
		Postal 14-740, 07000, Ciudad de M\'exico, M\'exico}
	
	\date{\today}
	
	\begin{abstract}
		A continuum approach to Quantum Chromodynamics (QCD), based upon Schwinger-Dyson (SD) and Bethe-Salpeter (BS) equations, is employed to provide a tightly constrained prediction for the $\gamma^{*} \gamma^{*} \rightarrow \{ \pi^0, \eta, \eta', \eta_c, \eta_b \}$ transition form factors (TFFs) and their  corresponding pole contribution to the hadronic light-by-light (HLbL) piece of the anomalous magnetic moment of the muon ($a_\mu$). This work relies on a practical and well-tested quark-photon vertex Ansatz approach to evaluate the TFFs for arbitrary space-like photon virtualities, in the impulse approximation. The numerical results are parametrized meticulously, ensuring a reliable evaluation of the HLbL contributions to $a_\mu$. We obtain: $a_{\mu}^{\pi^0-\textrm{pole}} = (6.14 \pm 0.21) \times 10^{-10}$, $a_{\mu}^{\eta-\textrm{pole}} = (1.47 \pm 0.19) \times 10^{-10}$, $a_{\mu}^{\eta'-\textrm{pole}} = (1.36 \pm 0.08) \times 10^{-10}$, yielding a total value of $a_{\mu}^{\pi^0+\eta+\eta'-\textrm{pole}} = (8.97 \pm 0.48) \times 10^{-10}$, compatible with contemporary determinations. Notably, we find that  $a_{\mu}^{\eta_c+\eta_b-\textrm{pole}} \approx a_{\mu}^{\eta_c-\textrm{pole}} = (0.09 \pm 0.01) \times 10^{-10}$, which might not be negligible once the percent precision in the computation of the light pseudoscalars is reached.
	\end{abstract}
	
	
	\maketitle
	
	\section{Introduction}
	
	More than half a century after the advent of the Standard Model (SM) of particle physics, it has
	successfully withstood a continuous barrage of innumerable experimental tests. Many of us are keenly interested 
	in high precision measurements of quantities which can be theoretically best calculated in order to zoom into 
	the very limits of this model, hunting for the possible discrepancies. Measurement and calculation of the muon 
	anomalous magnetic moment, $a_\mu=(g_\mu-2)/2$, provide precisely such battleground~\cite{Prades:2009tw, Jegerlehner:2009ry, Lindner:2016bgg}.
	The most recently reported value by the Brookhaven National Laboratory (BNL), $116592091(63)\times 10^{-11}$~\cite{Bennett:2006fi}
	shows a persistent $3.5$ standard deviations away from the SM prediction $116591823(43)\times 10^{-11}$~\cite{Tanabashi:2018oca}.
	Well deserved attention is currently being paid to this anomaly \footnote{There have been recent hints for an anomaly with opposite sign in $a_e$ at the $2.5\;\sigma$ level~\cite{Hanneke:2008tm, Parker:2018vye, Davoudiasl:2018fbb,Crivellin:2018qmi}.} due to the ongoing rigorous experimental endeavours to pin it down with increasing precision. The dedicated FNAL experiment will reach a fourfold improvement of the current statistical error within about two years from now~\cite{Grange:2015fou}. Later on, J-PARC also plans to achieve a comparable accuracy~\cite{Saito:2012zz}. If this deviation does not wither away, it would be highly desirable to reduce the SM calculational uncertainty as much as possible to be able to associate the {\em discrepancy} with possible new physics. What stand on the way are the hadronic contributions which are hard to tame and severely restrain our
	efforts to make predictions with the desired exactitude.
	
	The SM prediction includes quantum electrodynamics (QED) corrections up to five loops~\cite{Aoyama:2012wk, Aoyama:2014sxa, Aoyama:2017uqe}, two-loop (and leading-log three-loop) electroweak ones~\cite{Czarnecki:2002nt, Gnendiger:2013pva} and hadronic contributions, the latter saturating the error of the SM precision quoted above. These are divided into hadronic vacuum polarization and hadronic light-by-light (HLbL) contributions. While the former could be related to data already in 1961~\cite{Bouchiat:1957zz}, a similar data-driven extraction is not yet possible for the HLbL piece, although a dedicated effort~\cite{Colangelo:2014dfa, Colangelo:2014pva, Pauk:2014rfa, Colangelo:2015ama, Nyffeler:2016gnb, Danilkin:2016hnh, Colangelo:2017qdm, Colangelo:2017fiz, Hagelstein:2017obr} has made remarkable advances towards reaching this goal in the near future.
	
	The most recent evaluations of the hadronic vacuum polarization contribution to $a_\mu$~\cite{Jegerlehner:2017lbd, Davier:2017zfy, Keshavarzi:2018mgv, Colangelo:2018mtw, Hoferichter:2019gzf} have reduced its error, reaching the same level of uncertainty as the HLbL contribution. The latter must be diminished to fully benefit from the very precise forthcoming measurements at FNAL and J-PARC~\footnote{Subleading hadronic corrections are known at the required precision already \cite{Kurz:2014wya, Colangelo:2014qya}.}. The contributions of the lightest pseudoscalar mesons saturate $a_\mu^{\textrm{HLbL}}$~\footnote{This is, however, not understood from first principles \cite{deRafael:1993za, Knecht:2001qg}. See Refs. \cite{Knecht:2018sci} and \cite{Bijnens:2016hgx, Roig:2019reh} for recent approaches to compute scalar and axial-vector meson contributions, respectively.}, among which, the $\pi^0$-pole piece dominates~\cite{Hayakawa:1997rq,Blokland:2001pb, Bijnens:2001cq, Knecht:2001qf, Melnikov:2003xd, Erler:2006vu, Hong:2009zw, Goecke:2010if, Cappiello:2010uy, Kampf:2011ty, Dorokhov:2011zf, Masjuan:2012qn, Masjuan:2012wy, Roig:2014uja,Bijnens:2016hgx, Blum:2016lnc, Blum:2017cer, Masjuan:2017tvw, Guevara:2018rhj, Meyer:2018til}.

	In this paper, we compute the HLbL contributions coming from the light neutral pseudoscalar transition to two photons (and the first ever estimation for $\eta_c$ and $\eta_b$) to the anomalous magnetic moment of the muon. We follow a novel 
	Schwinger-Dyson and Bethe-Salpeter equations (SDEs, BSEs) approach to compute $\gamma \gamma^* \to M$~\cite{Raya:2015gva,Raya:2016yuj,Ding:2018xwy} transition form factor for arbitrarily large space-like momentum for the first time, in a unique framework with a direct connection to quantum chromodynamics (QCD). Such an approach is known to unify those form factors with their corresponding valence quark distribution amplitudes~\cite{Chang:2013pq,Segovia:2013eca,Ding:2015rkn}, charged pion and kaon form factors~\cite{Chang:2013nia,Gao:2017mmp,Chen:2018rwz}, their parton distribution functions~\cite{Chen:2016sno,Ding:2019lwe,Ding:2019qlr} and a wide range of other hadronic properties (masses, decay constants, etc.)~\cite{Chang:2011ei,Qin:2019hgk,Aguilar:2019teb}. We extend the SDE-BSE treatment of Refs.~\cite{Raya:2015gva,Raya:2016yuj,Ding:2018xwy} to account for arbitrary space-like virtualities of both photons. 
	
	Different plausible parametrizations of the numerical data are discussed. In particular, the flaws and strengths of \emph{Vector-Meson} and \emph{Lowest-Meson Dominance} (VMD, LMD) parametrizations~\cite{Knecht:2001qf} as well as \emph{Canterbury Approximants} (CAs)~\cite{Chisholm} are analyzed. In the context of $a_\mu^{\textrm{HLbL}}$, the latter were presented in Ref.~\cite{Masjuan:2017tvw} and also in a recent, but different, SDE-BSE approach~\cite{Eichmann:2019tjk}. As explained therein and below, we find the CAs parametrization more adequate.
	
	This article is organized as follows. In Sec.~\ref{DSEback}, we introduce the basics of our extension of Refs.~\cite{Raya:2015gva,Raya:2016yuj,Ding:2018xwy} to the doubly-off shell (DoS) case of the TFFs. The proposed parametrizations of the numerical data are presented in Sec.~\ref{Pi0cont}, together with the implications of the low and high energy behavior of the TFFs and the corresponding constraints. This framework is then applied to $\pi^0$, $\eta$ and $\eta'$ cases. This section ends with the corresponding description of the heavy $\eta_c$ and $\eta_b$ mesons. In Sec.~\ref{Results} we discuss our results for $a_\mu^{\textrm{HLbL}}$. Based on this analysis, we present our conclusions in Sec.~\ref{Concl}.
	
	\section{SDE-BSE approach} \label{DSEback}
	
	The transition $\gamma^{*}\gamma^{*} \to M$ is described by a single form factor. In the impulse approximation~\cite{Raya:2015gva}, 
	\begin{eqnarray}
	\mathcal{T}_{\mu\nu}(Q_1,Q_2)&=& T_{\mu\nu}(Q_1,Q_2) + T_{\nu\mu}(Q_2,Q_1)\;,\\ \nonumber
	T_{\mu\nu}(Q_1,Q_2) &=& \frac{e^2}{4\pi^2 }\epsilon_{\mu \nu \alpha \beta} Q_{1\alpha} Q_{2\beta} G(Q_1^2,Q_1\cdot Q_2,Q_2^2) \\ \nonumber
	&=& \mathbf{e}_M^2 \; \text{tr}_{CD}\int_q i  \chi_{\mu}^f(q,q_1)\Gamma_{M}(q_1,q_2) \\
	&\times& S_f(q_2) i \Gamma_\nu^f(q_2,q)\;,\label{eq:defTFF}
	\end{eqnarray}
	where $Q_{1},\;Q_2$ are the momenta of the two photons and $(Q_1+Q_2)^2=P^2 = - m_M^2$ ($m_M$ is the mass of the pseudoscalar). The kinematic arrangement is $q_1=q+Q_1$, $q_2=q-Q_2$; with $q$ being the integration variable \footnote{For simplicity in the notation, we have defined $\int_q \equiv \int \frac{d^4q}{(2\pi)^4}$.}. In addition, $\mathbf{e}_M=e\;\mathbf{c}_M$ is a charge factor associated with the valence quarks of the given meson ($e$ is the charge of the positron)~\footnote{As explained in Ref.~\cite{Ding:2018xwy}, Eq.~\eqref{eq:defTFF} is modified to account for the flavour decomposition of the $\eta-\eta'$ systems.}. The other symbols carry their usual meanings:
	\begin{itemize}
		\item $S_f(p)=-i\gamma \cdot p \;\sigma_v(p^2) + \sigma_s(p^2)$ is the propagator of the $f$-flavoured quark. It is determined from its SDE (namely, the \emph{gap equation}):
		\begin{eqnarray}
		\nonumber
		S_f^{-1}(p) &=& \mathcal{Z}_{2} (-i \gamma \cdot p + m_f) + \Sigma_f(p) \;, \\
		\label{eq:prop2}
		\Sigma_f(p) &=& -\int_q [\mathcal{K}(q,p)]_{rs}^{tu} S_{sr}(q) \;, 
		\end{eqnarray}    
		where $\mathcal{K}(q,p)$ is the kernel of the gap equation and $\{r,s,t,u\}$ are color indices (not displayed when obvious). Quark propagator, and every other Green function involved in its SDE, are renormalized at the resolution scale of $\zeta = 2 \; \textrm{GeV}:= \zeta_2 $.
		
		\item $\Gamma_M(p;P)$ is the Bethe-Salpeter amplitude of the pseudoscalar meson $M$, obtained from its BSE:
		\begin{eqnarray}    
		\Gamma_{M}(p;P) &=& \int_q [\chi_{M}(q;P)]_{sr} \mathcal{M}_{rs}^{tu}(q,p;P) \;,
		\end{eqnarray}
		where $P$ is the total momentum of the bound state and $\chi_{M}(q;P)=S(q+\eta P)\Gamma_{M}(q;P)S(q-(1-\eta)P)$, $\eta \in [0,1]$. No physical observable depends on $\eta$, the definition of the relative momentum. 
		
		Here $\mathcal{M}(q,p;P)$ is the renormalized, fully amputated, two particle irreducible, quark-antiquark scattering kernel. It is related to $\mathcal{K}$ via the Axial-vector Ward-Takahashi identity~\cite{Chang:2009zb,Qin:2014vya}. 
		
		\item Finally, we have the amputated, $\Gamma_\mu(q_f,q_i)$, and unamputated, $\chi_\mu(q_f,q_i)$, quark-photon vertices (QPV). Those obey their own SDEs~\cite{Eichmann:2019tjk,Eichmann:2017wil}. 
	\end{itemize}
	
	In conjunction with Eq.~\eqref{eq:defTFF}, we employ the so-called Rainbow-Ladder truncation (RL), which is known to accurately describe the pseudoscalar mesons~\cite{Raya:2015gva,Raya:2016yuj,Ding:2018xwy,Chang:2013nia,Ding:2019qlr}. This entails: 
	\begin{eqnarray}
	\nonumber
	\mathcal{M}_{tu}^{rs}(q,p;P) &=& \mathcal{K}_{tu}^{rs}(q,p) \\
	&\equiv& -\frac{4}{3} \mathcal{G}(k^2) D_{\mu\nu}^0(k) [\gamma_\mu]_{ts} [\gamma_\nu]_{ru}\;,
	\end{eqnarray}
	where $k=p-q$, $D_{\mu\nu}^0(k)$ is the tree-level gluon propagator in the Landau gauge and $\mathcal{G}(k^2)$ is an effective dressing function. We employ the well-known Qin-Chang interaction~\cite{Qin:2011dd}, compatible with our modern understanding of the gluon propagator~\cite{Binosi:2016nme,Huber:2017txg,Rodriguez-Quintero:2018wma, Cui:2019dwv}: its dressing function saturates in the infrared and monotonically decreases as the momentum increases and recovers the perturbative QCD running coupling in the ultraviolet. With the interaction strength ($\omega D = m_G^3$) fixed, all physical observables are practically insensitive to variations of $\omega \in (0.4,0.6)$ GeV~\cite{Qin:2011dd,Chen:2018rwz}. Moreover, sensible variations of $m_G$ do not alter the output observables significantly either, a fact that is illustrated through the computation of the pion mass and decay constant with two parameter sets (all these details are taken into account in our final result). Here on, we shall employ isospin symmetry $m_u = m_d := m_l$. Typically, $m_G$ and the current quark masses are fixed such that ground-state masses and decay constants are properly reproduced~\cite{Qin:2011dd,Eichmann:2019tjk}. Thus, for our first set of parameters (RL-I), we employ $\{m_\pi$, $m_K$, $f_\pi$, $f_K\}$ as benchmarks; for the second set (RL-II), we use $\{m_\pi$, $m_{\eta_c}$, $m_{\eta_b}$, $f_\pi\}$ instead. Precise input parameters, computed masses and decay constants are given in Table~\ref{Tab:RLparams2}.

	\subsection{Quark-photon vertex}
	 In principle, the QPV can be obtained from its inhomogeneous BS equation. This process automatically incorporates vector meson poles~\cite{Maris:1999bh,Eichmann:2017wil} in the vertex and guarantees the preservation of the Abelian anomaly~\cite{Adler:2004ih, Bell:1969ts,Adler:1969gk}. However, it also limits the domain in which a direct evaluation of the TFFs is possible~\cite{Maris:2002mz,Chen:2016bpj,Eichmann:2017wil,Eichmann:2019tjk}. Thus we follow an alternative route. We introduce a reliable QPV Ansatz based upon gauge covariance properties and multiplication renormalizability of the massless fermion propagator. In conjunction with this approach, we incorporate the non-Abelian anomaly at the level of the BSE. This approach sets apart our work (and our previous ones~\cite{Raya:2015gva,Raya:2016yuj,Ding:2018xwy}) from the recent SDE approach of~\cite{Eichmann:2019tjk,Eichmann:2017wil}.
		
		A kindred version of the vertex Ansatz we employ was first introduced in~\cite{Chang:2013nia}, for the calculation of the pion elastic form factor (EFF) and subsequently adapted in~\cite{Raya:2015gva} for the TFFs. The QPV is expressed completely via the functions which characterize the dressed quark propagator ($q = k_f - k_i$, $\bar{\mathbf{s}} = 1 - \mathbf{s}$):
	\begin{eqnarray}
	\nonumber
	\chi_\mu(k_f,k_i)&=& \gamma_\mu \Delta_{k^2 \sigma_V} \\
	\nonumber
	&+& [\mathbf{s} \gamma\cdot k_f \gamma_\mu \gamma \cdot k_i + \bar{\mathbf{s}}\gamma\cdot k_i \gamma_\mu \gamma \cdot k_f]\Delta_{\sigma_V}\nonumber\\
	\nonumber
	&+&[\mathbf{s}(\gamma\cdot k_f \gamma_\mu + \gamma_\mu \gamma \cdot k_i)\\
	&+&\bar{\mathbf{s}}(\gamma\cdot k_i \gamma_\mu + \gamma_\mu \gamma \cdot k_f)]i\Delta_{\sigma_S}\;,
	\label{eq:vertex}
	\end{eqnarray}
	where $\Delta_F=[F(k_f^2)-F(k_i^2)]/(k_f^2-k_i^2)$, $\bar{\mathbf{s}}=1-\mathbf{s}$. Up to transverse pieces associated with $\mathbf{s}$, $\chi_\mu(k_f,k_i)$ and $S(k_f)\Gamma_\mu(k_f,k_i)S(k_i)$ are equivalent. Longitudinal pieces alone do not recover the Abelian anomaly, since it turns out impossible to simultaneously conserve the vector and axial-vector currents associated with Eq.~\eqref{eq:defTFF}. Thus a momentum redistribution factor is introduced:
	\begin{equation}
	\label{eq:TTs}
	\mathbf{s} = \mathbf{s}_f\; \textrm{exp}\left[-\left(\sqrt{Q_1^2/4 + m_M^2}-m_M\right)/M_E^f\right]\;,
	\end{equation}
	where $M_E^f = \{ p | p^2 = M_f^2(p^2),\; p^2\textgreater0 \}$ and $M_f(p^2)$ is the quark's mass function. As $\mathbf{s}$ is exponentially suppressed, it does not affect the large $Q^2$ behavior of the TFFs. To account for $Q_2^2\neq 0$, the simplest symmetrization corresponds to the replacement $Q_1^2 \to Q_1^2 + Q_2^2$, which clearly recovers all the limits \footnote{Additional subtleties appear in the large-$Q^2$ regime, concerning the QCD evolution of the TFFs. This discussion will be addressed elsewhere, since it is not relevant to the HLbL computations, fully determined by the low-$Q^2$ region. }. The way the specific values of $\mathbf{s}_f$ are established is addressed in Sections~\ref{limits} and~\ref{Results}. The following pattern is observed:
		\begin{eqnarray}
		\label{eq:valuessf}
		\mathbf{s}_{l}\simeq 1.91\;\textgreater\;\mathbf{s}_c \simeq 0.78 \; \textgreater \; \mathbf{s}_b \simeq 0.23\;.
		\end{eqnarray}
	Given the proposed form in Eq.~\eqref{eq:vertex}, the QPV is determined by the quark propagator dressing functions. Consequently, the TFFs are fully expressed in terms of quark propagators and BS amplitudes, obtained in the RL truncation. We then employ perturbation theory integral representations (PTIRs) for those objects, as previously done in the calculation of the pion distribution amplitude~\cite{Chang:2013pq} and its EFF~\cite{Chang:2013nia}. The particular representations were presented in~\cite{Chang:2013pq} and discussed herein in the appendix. PTIRs allow us to write Eq.~\eqref{eq:defTFF} in terms of objects which have $q$-quadratic forms in the denominator.  Thus, after introducing Feynman parametrization and a suitable change of variables, the 4-momentum integrals can be evaluated analytically. Subsequently, integrations over the Feynman parameters and the spectral density are performed numerically. The complete calculations basically require a series of perturbation-theory-like integrals. This expedites the computation considerably and allows a direct evaluation of the TFFs in the whole domain of space-like momenta.  We thus managed, for the very first time, to compute the $\gamma^* \gamma^* \to$ {\em All} lowest-lying neutral pseudoscalars TFFs in this domain.

	In using the proposed QPV Ansatz, we overcame the inconvenience stemming from solving its BS equation and it expedited the computation of the TFFs. Despite lacking explicit non-analytic structures, associated with vector meson poles in the time-like region, the effects in the space-like region are appropriately reproduced (later discussed in connection with the charge radius). Thus we expect our approach to be valid by essentially maintaining
		the key quantitative details. Our Ansatz follows from using the gauge technique~\cite{Delbourgo:1977jc}. Thus it satisfies the longitudinal Ward-Green-Takahashi identity (WGTI)~\cite{Ward:1950xp, Green:1953te, Takahashi:1957xn}, is free of kinematic singularities, reduces to the bare vertex in the free-field limit, and has the same Poincar\'e transformation properties as the bare vertex.
	
	\subsection{The $\eta-\eta'$ case} \label{eta-etap}
	Our dealing with the $\eta-\eta'$ mesons is now discussed. First, we use a flavour basis to rewrite the BS amplitudes as follows:
	\begin{eqnarray}
	\Gamma_{\eta,\eta'}(k;P) &=& \textrm{diag}(1,1,0) \Gamma_{\eta,\eta'}^l(k;P)\nonumber\\
	&+&\textrm{diag}(0,0,\sqrt{2}) \Gamma_{\eta,\eta'}^s(k;P)\;,
	\end{eqnarray}
	where we keep using the isospin symmetric limit, such that $l = u,d$. The RL kernel by itself does not produce any mixing between the pure $l\bar{l}$ and $s\bar{s}$ states. Thus, the Bethe-Salpeter kernel is improved by including the non-Abelian anomaly kernel (see Refs.~\cite{Bhagwat:2007ha, Ding:2018xwy}):
	\begin{eqnarray}
	\nonumber
	\mathcal{M}_{tu}^{rs}(q,p;P) &=& \mathcal{K}_{tu}^{rs}(q,p) + \mathcal{A}_{tu}^{rs}(q,p;P)\;, \\ \nonumber
	\mathcal{A}_{tu}^{rs}(q,p;P) &\equiv& -\mathcal{G}_A(k^2)\huge( \sin^2{\theta_\xi} [\mathbf{r} \gamma_5]_{rs}  [\mathbf{r} \gamma_5]_{tu} \\
	&+& \frac{1}{\chi_l^2} \cos^2{\theta_\xi} [\mathbf{r} \gamma_5 \gamma \cdot P]_{rs}  [\mathbf{r} \gamma_5 \gamma \cdot P]_{tu} \huge)\;, \quad \label{eq:anomaly}
	\end{eqnarray}
	with $\chi_l = M_l(0)$ and $\theta_A$ controlling the relative strength between the $\gamma_5$ and $\gamma_5 \gamma \cdot P$ terms; $\mathbf{r}=$diag$(1,1,\nu_R)$, where $\nu_R = M_l(0)/M_s(0) = 0.57$. It models a dependence on $U(3)$ flavour-symmetry breaking arising from the dressed-quark lines which complete a `U-turn' in the hairpin diagram (see Fig.~1 in Ref.~\cite{Ding:2018xwy}). The strength of the anomaly is controlled by
	\begin{equation}
	\mathcal{G}_A(k^2) = \frac{8 \pi^2}{\omega_\xi^4} D_\xi\; \textrm{exp}[-k^2/\omega_\xi^2]\;.
	\label{eq:anomaly2}
	\end{equation}
	Here $\omega_\xi$ and $D_\xi$ provide a momentum dependence for the anomaly kernel, as a generalization to that introduced in Ref.~\cite{Bhagwat:2007ha}. The set of Dirac covariants which describe Eq.~\eqref{eq:anomaly} can be inferred from the axial-vector WGTI~\cite{Bhagwat:2007ha}; we keep those which dominate. The rest of the pieces are not determined by the WGTI, but they can be driven by phenomenology~\cite{Ding:2018xwy}. Since the RL truncation does not produce any mixing by itself, it is natural to require more input to describe the $\eta-\eta'$ system. In particular, given the anomaly kernel of Eqs.~\eqref{eq:anomaly}-\eqref{eq:anomaly2}, we fix $D_\xi,\;\omega_\xi$ and $\cos^2 \theta_\xi$ to provide a fair description of $m_{\eta,\eta'}$ and $f_{\eta,\eta'}^l$~\footnote{Thus, in addition to the usal setting of the free RL parameters~\cite{Qin:2011dd,Eichmann:2019tjk}, 3 more are introduced to obtain 6 new observables.}. More weight is given to the masses, which are better constrained empirically. Input and output values, together with the RL counterpart, are listed in Table~\ref{Tab:RLparams2}. As a reference, if a single mixing angle scheme (and a pair of ideal decay constants) is assumed, our results yield $f^l \approx 1.08 f_\pi $, $f^s \approx 1.49 f_\pi$ and $\phi_{\eta\eta'} = 42.8\degree$.  Moreover, note that $D_\xi = 0$ turns off the non-Abelian anomaly and produces an ideal mixing with pure $l \bar{l}$ and $s \bar{s}$ states, which implies  $m_\eta=m_\pi=0.135$ GeV and $m_{\eta'}=m_{ss}=0.698$ GeV ($f^l:=f_\pi = 0.093$ GeV, $f^s:=f_{ss}=0.134$ GeV).

	\begin{table*}[htbp]
	\caption{\label{Tab:RLparams2} RL parameters (left and central panels) are fixed to produce the ground-state masses and decay constants. We restrain ourselves to $\omega=0.5$ GeV, the midpoint of the domain of insensitivity~\cite{Qin:2011dd}. The $\eta-\eta'$ values follow from RL-I parameters plus the anomaly kernel (Eq.\eqref{eq:anomaly}) inputs given in the right panel. Experimental (Exp.) PDG values are taken from~\cite{Tanabashi:2018oca}, ($^*$) lattice QCD results from~\cite{Koponen:2017fvm} and ($^\dagger$) phenomenological reference numbers from~\cite{Ding:2018xwy}; here $m_\pi$ and $m_K$ correspond to the average of the neutral and charged mesons. The mass units are in GeV.}

	\begin{center}
	\begin{tabular}{@{\extracolsep{0.3 cm}}ccc}
	
	\begin{tabular}{@{\extracolsep{0.2 cm}}l|l||l|l|l}
		\hline \hline
\multicolumn{1}{l}{RL-I} & & &  Herein & Exp. \\
	\hline
	$m_G$ & $0.80$ & $m_\pi$ & $0.135$ & $0.137$ \\
	$m_l$ & $0.0051$ & $m_K$ & $0.496$ & $0.496$ \\
	$m_s$ & $0.125$ & $m_{ss}$ & $0.698$ & $0.689^*$ \\
	 &  & $f_\pi$ & $0.093$ & $0.093$ \\
	 &  & $f_K$ & $0.112$ & $0.111$ \\
	&  & $f_{ss}$ & $0.134$ & $0.128^*$ \\
		\hline	
	\end{tabular} &

	\begin{tabular}{@{\extracolsep{0.2 cm}}l|l||l|l|l}
	\hline \hline
	\multicolumn{1}{l}{RL-II} & & & Herein & Exp. \\
	\hline
	$m_G$ & $0.87$ & $m_\pi$ & $0.138$ & $0.137$ \\
	$m_l$ & $0.0042$ & $m_{\eta_c}$ & $2.981$ & $2.984$ \\
	$m_c$ & $1.21$ & $m_{\eta_b}$ & $9.392$ & $9.399$ \\
	$m_b$ & $4.19$ & $f_\pi$ & $0.093$ & $0.093$ \\
	&  & $f_{\eta_c}$ & $0.262$ & $0.237$ \\
	&  & $f_{\eta_b}$ & $0.543$ & $--$ \\
	\hline	
\end{tabular} 	&

\begin{tabular}{@{\extracolsep{0.2 cm}}l|l||l|l|l}
	\hline \hline
	\multicolumn{1}{c}{A. Kernel} & & & Herein & Exp. \\
	\hline
	$\sqrt{D_\xi}$ & $0.32$ & $m_\eta$ & $0.560$ & $0.548$ \\
	$\omega_\xi$ & $0.30$ & $m_{\eta'}$ & $0.960$ & $0.958$ \\
	$\cos^2 \theta_\xi$& $0.80$ & $f_{\eta}^l$ & $0.072$ & $0.090^{\dagger}$ \\
	&  & $-f_{\eta}^s$ & $0.092$ & $0.093^{\dagger}$ \\
	&  & $f_{\eta'}^l$ & $0.070$ & $0.073^{\dagger}$ \\
	&  & $f_{\eta'}^s$ & $0.101$ & $0.094^{\dagger}$ \\
	\hline	
\end{tabular} 
	
	\end{tabular}
	\end{center}
\end{table*}	
	
	\subsection{Kinematical limits} \label{limits}
	We now turn our attention to the transition form factors, which we define in the standard way (see e.g. Ref.~\cite{Lepage:1980fj}), focusing  on large $Q^2$ behavior 
	to start with. It is well known that above a certain large scale $\tilde{Q}_0^2 \textgreater \Lambda_{\textrm{QCD}}^2$, the TFFs take the form~\cite{Brodsky:1973kr,Lepage:1980fj,Maris:2002mz}:
	\begin{eqnarray}
	Q^2 F_M(Q^2,0) &\to& 2 f_M \mathbf{c}_M^2 \int_0^1 dx \; \frac{\phi_M^q(x;Q^2)}{x}  \;,\\
	Q^2 F_M(Q^2,Q^2) &\to& 2 f_M \mathbf{c}_M^2 \int_0^1 dx \; \phi_M^q(x;Q^2)\;, 
	\end{eqnarray}
	where $Q^2\textgreater\tilde{Q}_0^2\;$ and $\phi_M^q(x;Q^2)$ is the $q$-flavour valence quark distribution amplitude of meson $M$. For notational convenience, and in order to match experimental normalization, the TFFs have been rescaled as $F_M(Q_1^2,Q_2^2) \to F_M(Q_1^2,Q_2^2)/(2\pi^2)$. This normalization will be employed from this point onward. In the asymptotic domain, $Q^2 \rightarrow \infty$, where the conformal limit (CL) is valid, one arrives at:
	\begin{eqnarray}
	\phi_M^q(x;Q^2) & \stackrel{Q^2 \to \infty}{\rightarrow } & \phi_{\textrm{CL}}(x) = 6x(1-x)\;,
	\end{eqnarray}
	from which the corresponding limits of the SoS and equally off-shell (EoS) TFFs are obtained:
	\begin{eqnarray}
	\label{eq:BLlimit1}
	Q^2 F_M(Q^2,0) & \stackrel{Q^2 \to \infty}{\rightarrow } & 6 f_M \mathbf{c}_M^2\; \equiv F_M^{\infty}, \\
	\label{eq:BLlimit2}
	Q^2 F_M(Q^2,Q^2) & \stackrel{Q^2 \to \infty}{\rightarrow } & 2 f_M \mathbf{c}_M^2\; = \frac{F_M^{\infty}}{3}.
	\end{eqnarray}
	For the pion, $\mathbf{c}^2_\pi=e^2(4-1)/9$, thus arriving at the well-known limit $F_M^\infty = 2 f_\pi$~\cite{Brodsky:1973kr,Lepage:1980fj}. To account for the flavour structure of the $\eta-\eta'$ systems, Eq.~\eqref{eq:BLlimit1} is modified as
	\begin{eqnarray}
	Q^2 F_{\eta,\eta'}(Q^2,0) &\stackrel{Q^2 \to \infty}{\rightarrow }& 6[c_l f_{\eta,\eta'}^l(Q^2) + c_s f_{\eta,\eta'}^s(Q^2)] \quad \nonumber \\
	&=& 2[c_8 f_{\eta,\eta'}^8 + c_0 f_{\eta,\eta'}^0(Q^2)]\,, \label{eq:BLlimit3}
	\end{eqnarray}
	where $c_l = 5/9$, $c_s = \sqrt{2}/9$, $c_8=1/\sqrt{3}$, $c_0=\sqrt{2/3}$; and $f_{\eta,\eta'}^{8,0}$ are the decay constants in the octet-singlet basis~\cite{Leutwyler:1997yr}. Owing to the non-Abelian anomaly, the singlet decay constant, $f_{\eta,\eta'}^0$, and thus $f^{l,s}_{\eta,\eta'}$, exhibits scale dependence~\cite{Agaev:2014wna}. Moreover, although the RL gives the correct power laws as perturbative QCD, it fails to produce the correct anomalous dimensions. This is readily solved by a proper evolution of the BS wave function.  The way QCD evolution is implemented in our calculations is detailed in Refs.~\cite{Raya:2015gva,Raya:2016yuj,Ding:2018xwy}; this process entails that the correct limits of Eqs.~\eqref{eq:BLlimit1}-\eqref{eq:BLlimit2} are numerically reproduced. 
	
	In the opposite kinematical limit of $Q^2$, the Abelian anomaly dictates the strength of $F_M(0,0)$ for the Goldstone modes.  In the chiral limit, this entails:
	\begin{eqnarray}
	\label{eq:AA1}
	F_M(0,0) = \frac{1}{4\pi^2 f_\pi^0}\;,
	\end{eqnarray}
	where the  index `$0$' denotes chiral limit value.  The non-masslessness of the $\pi^0$ and $\eta$ mesons produces slight deviations from the above result~\cite{Tanabashi:2018oca,Danilkin:2019mhd}. Supplemented by our value $f_\pi^0 = 0.092$ GeV, Eq.~\eqref{eq:AA1} can be employed to fix $\mathbf{s}_{0}=1.91\simeq \mathbf{s}_l$ in the QPV, Eq.~\eqref{eq:vertex}.
		
		On the other hand, the TFFs at $Q^2=0$ are also related to their corresponding decay widths, $\Gamma(\gamma \gamma \to M):=\Gamma_M^{\gamma\gamma}$, via the equation:
		\begin{eqnarray}
		\label{eq:FvsDecay}
		F_M(0,0) = \sqrt{\frac{4\Gamma_M^{\gamma\gamma}}{\pi \alpha_{em}^2 m_M^3}}\;,
		\end{eqnarray}
		with $\alpha_{em}=e^2/(4\pi)$, the electromagnetic coupling constant. From the computed masses and decay constants in Table~\ref{Tab:RLparams2}, one can readily infer the corresponding decay widths for the $\eta_{c}$ and $\eta_b$ mesons~\cite{Raya:2016yuj}:
		\begin{equation}
		\Gamma_{\eta_{c,b}}^{\gamma\gamma} = \frac{8 \pi \alpha_{em}^2 \mathbf{c}_{\eta_{c,b}}^4 f_{\eta_{c,b}}^2}{m_{\eta_{c,b}}}\;,\textrm{with } \mathbf{c}_{\eta_{c,b}}=2/3,1/3\;.
		\end{equation}
		This yields to the values:
		\begin{equation}
		\Gamma_{\eta_c}^{\gamma\gamma}=6.1\;\textrm{keV}\;,\;\Gamma_{\eta_b}^{\gamma\gamma}=0.52\;\textrm{keV}\;,
		\end{equation}
		such that $\mathbf{s}_c = 0.78$ and $\mathbf{s}_b=0.23$ in order to hold Eq.~\eqref{eq:FvsDecay} true~\footnote{Experimentally, $\Gamma_{\eta_{c}}^{\gamma\gamma}=5.0(4) $ keV. Nothing is gained for the TFF if $\mathbf{s}_c$ is fixed to reproduce that value,~\cite{Raya:2016yuj}, and the corresponding contribution to $a_\mu$ would be contained within our final error estimate.}. Current algebra is adapted to obtain the analogous for the $\eta-\eta'$ case~\cite{Ding:2018xwy}:
		\begin{equation}
		\Gamma_{\eta,\eta'}^{\gamma\gamma} = \frac{9 \alpha_{em}^2 m_{\eta,\eta'}^3}{64 \pi^3} \left[c_l \frac{f_{\eta,\eta'}^l}{(f^l)^2}+c_s \frac{f_{\eta,\eta'}^s}{(f^s)^2} \right]^2\;.
		\end{equation}
		Thus, from the values of Table~\ref{Tab:RLparams2}, one gets:
		\begin{equation}
		\label{eq:DWetaetap}
		\Gamma_{\eta}^{\gamma\gamma}=0.42\;\textrm{keV}\;,\;\Gamma_{\eta'}^{\gamma\gamma}=4.66\;\textrm{keV}\;,
		\end{equation}
		predictions which are commensurate with empirical determinations, respectively~\cite{Tanabashi:2018oca}:  $0.516(22)$ keV, $4.35(36)$ keV. The results of~\eqref{eq:DWetaetap} fix $\mathbf{s}_s=0.48$ and demand a reduction of $\mathbf{s}_l=1.91 \to 1.21$, for the $\eta-\eta'$ case. This happens for two reasons: since we give more weight to the correct description of masses, our best set of parameters in Table~\ref{Tab:RLparams2} underestimates the value of $f_\eta^l$, as compared to phenomenology. Secondly, a key difference of $\eta-\eta'$, with respect to $\pi^0,\;\eta_c\;,\eta_b$, is the presence of the  non-Abelian  anomaly,  which conceivably generates corrections to Eq.~\eqref{eq:defTFF} at infrared momenta.  This issue will be addressed elsewhere. Nonetheless, we can estimate the potential impact of our model inputs by following the criteria explained in Section~\ref{Results}. 
	
	A comparison of our results for $F_M(0,0)$ to the corresponding measurements is given in Table~\ref{Tab:widths}. Basically, the error bars are obtained by varying the strength of the transverse terms in the QPV ($\mathbf{s}_f$), as well as the computed masses. This process does not alter any of the conclusions presented in Refs.~\cite{Raya:2015gva,Raya:2016yuj,Ding:2018xwy}, nor the agreement those results have with the empirical data; instead, it allows us to provide an error estimate in a quantity that could be sensitive to small variations of the inputs, such as $a_\mu$.
	
	\begin{table}[htbp]
		\caption{\label{Tab:widths} Inferred values of $F_M(0,0)$, considering the error estimate criteria of Section~\ref{Results}. Results given in GeV$^{-1}$. }
		\begin{center}
			\begin{tabular}{@{\extracolsep{0.0 cm}}lll}
				\hline \hline
				Meson & This Work & Experiment \cite{Tanabashi:2018oca} \\
				\hline
				$\pi^0$ & $0.2753\;(31)$ & $0.2725\;(29)$ \\
				$\eta$ & $0.2562\;(170)$ & $0.2736\;(60)$ \\
				$\eta'$ & $0.3495\;(60)$ & $0.3412\;(76)$ \\
				$\eta_c$ & $0.0705\;(40)$ & $0.0678\;(30)$ \\
				$\eta_b$ & $0.0038\;(2)$ & $--$ \\
				\hline \hline
			\end{tabular}
		\end{center}
	\end{table}
	
	Notice that, while the $F(0,0)$ values of $\pi^0$, $\eta'$ and $\eta_c$ show an accurate match with the empirically inferred results, the $\eta$ is underestimated and produces a larger error bar. A similar pattern has been observed for the charge radii ($r_M$), which is essentially the slope at $Q^2=0$. While the $\pi$ and $\eta_c$ charge radii are obtained with desired accuracy~\cite{Raya:2015gva,Raya:2016yuj} (this also occurs with the pion EFF~\cite{Chang:2013nia,Chen:2018rwz}), and the corresponding comparison with the experiment is within the $1.5\%$ level~\cite{Tanabashi:2018oca}, the $\eta-\eta'$ system suffers from a larger uncertainty~\cite{Ding:2018xwy}. This is attributed mostly to the presence of the non-Abelian anomaly and the failure of Eq.~\eqref{eq:defTFF} to incorporate beyond RL effects.  Firstly, it is worth mentioning that although our TFF lacks (dynamical) poles in the time-like region, the space-like behavior at low-$Q^2$ is compatible with VMD, $F_M(Q^2,0)\sim(Q^2+m_V^2)^{-1}$, as can be read from Figure~\ref{fig:pionTFF}. Second, our vertex Ansatz is further validated by the neat agreement with Ref.~\cite{Maris:1999bh}, in which the connection between the QPV and the pion charge radius is clearly established. While our analysis yields $r_{\pi} = 0.675(9)$, the most complete result in~\cite{Maris:1999bh} gives  $r_{\pi} = 0.678$ fm. Finally, our $\pi^0,\;\eta_c,\;\eta_b,$ predictions (obtained with QPV Ansatz and PTIRs) have been proven entirely compatible with those approaches that solve the vertex BSE instead and perform a direct calculation~\cite{Maris:2002mz,Chen:2016bpj}. 
	
	Given the set of SDE-BSE inputs and results, in the next section we discuss the parametrizations for the obtained numerical data.

	\section{Parametrizations and constraints} \label{Pi0cont}
	Regardless of the approach one takes to compute TFFs, it is highly convenient to look for certain types of theory-driven parametrizations for those form factors, such that the corresponding integrals of $a_\mu^{\textrm{HLbL}}$ can be computed with relative ease and with a minimum error following standard methods~\cite{Knecht:2001qf,Roig:2014uja}.
	
	Besides accurately fitting the numerical data, we look for parametrizations that reproduce the low and high $Q^2$ constraints to the fullest extent possible. We now discuss VMD, LMD and CAs parametrizations.
	
	\subsection{LMD parametrizations: flaws and strengths} 
	
	For considerable time, VMD and LMD type of parametrizations have been quite popular. Among other attractive aspects, they allow us to rewrite the $a_\mu^{\textrm{HLbL}}$-related integrals in such a way that there is no dependence on $Q_1 \cdot Q_2$~\cite{Jegerlehner:2009ry,Knecht:2001qf}. 
	
	Nevertheless, VMD and LMD fail in reproducing the large $Q^2$ limits, yielding an incorrect power law in both the SoS (Eq.~\eqref{eq:BLlimit1}) or EoS (Eq.~\eqref{eq:BLlimit2}) cases. Extensions of LMD that include one or more additional vector mesons (LMD+V or LMD+V+V') can potentially fulfill such requirements. Due to a higher number of parameters, such attempts can provide a more reliable fit in a larger domain of momenta.
	
	In general terms, with $x=Q_1^2$ and $y=Q_2^2$, one can define the LMD+$V_{N-1}$ parametrizations as follows:
	\begin{eqnarray}
	P(x,y) &=& \sum_{a,b} c_{a,b} (x+y)^a (xy)^b\;, \\
	& & \{a,b \in \mathbb{N}|0 \leq a+b\leq N  \} \nonumber \\
	R(x,y) &=& \prod_{i=0}^{N-1} (x+M_{V_{i}}^2)(y+M_{V_{i}}^2)\;,\\
	F_M(x,y) &=& \frac{P(x,y)}{R(x,y)}\;,
	\end{eqnarray}
	where $c_{a,b}$ and $M_{V_i}$ are the fitting parameters ($M_{V_i}$, can be related to the ground state vector mesons and its excitations). $N=1$ reproduces the usual LMD parametrization. Demanding $c_{N,0} = c_{0,N}=0$, both $x F_M(x,0)$ and $x F_M(x,x)$ tend to a constant as $x$ grows. Thus one can impose the asymptotic constraints of Eqs. (\ref{eq:BLlimit1})-(\ref{eq:BLlimit2}) and get
	\begin{equation}
	c_{N-1,0} = F_M^\infty \left(\prod_{i=0}^{N-1} M_{V_i}^2 \right)\;,\;\;
	c_{1,N-1}= \frac{1}{6} F_M^\infty.
	\end{equation}
	However, for any finite $y_0\neq 0$, $x F_M(x,y_0)$ diverges linearly with $c_{N-1,1}+y_0\; c_{N-2,2}$ as $x\to \infty$.
	Consequently, the asymptotic limits  cannot be recovered for arbitrary $y_0$ and the accuracy of the fit is compromised as $x$ increases, which makes this parametrization unsatisfactory. 

	\subsection{Canterbury Approximants}\label{CA}
	A more convenient approach to the problem at hand is through the so called \emph{Canterbury Approximants}~\cite{Chisholm}, which have been recently employed to evaluate $a_\mu^{\textrm{HLbL}}$~\cite{Masjuan:2017tvw,Eichmann:2019tjk}. The latter reference follows another SDE treatment to evaluate the pole contributions of $\pi^0,\; \eta, \;\eta'$ to $a_\mu^{\textrm{HLbL}}$.
	
	We explore this alternative to parametrize our numerical solutions and calculate the respective contributions to $a_\mu^{\textrm{HLbL}}$. Consider a function $f(x,y)$ symmetric in its variables and with a known series expansion
	\begin{eqnarray*}
		\label{eq:expCant}
		f(x,y) = \sum_{i,j} c_{i,j} x^i y^j \;,\;(c_{i,j}=c_{j,i})\;.
	\end{eqnarray*}
	CAs are defined as rational functions constructed out of such polynomials $P_N(x,y)$ and $R_M(x,y)$:
	\begin{equation}
	C_M^N(x,y) = \frac{P_N(x,y)}{R_M(x,y)} = \frac{\sum_{i,j=0}^N a_{ij} x^i y^j}{\sum_{i,j=0}^M b_{ij} x^i y^j}\;,
	\end{equation}
	whose coefficients $a_{ij}, b_{ij}$ fulfill the mathematical rules explained in detail in Ref.~\cite{Masjuan:2017tvw}. 
	We shall employ a certain $C_{2}^1(x,y)$ such that the TFFs can be written as:
	\begin{eqnarray}
	P(x,y) &=& a_{00}+a_{10}(x+y)+a_{01}(xy)\;,\\
	\nonumber
	R(x,y) &=& 1+b_{10}(x+y)+b_{01}(xy)+b_{11}(x+y)(xy)\\
	&+&b_{20}(x^2+y^2) \;,\\
	F_M(x,y) &=& \frac{P(x,y)}{R(x,y)}\;. \label{eq:CAsdef}
	\end{eqnarray}
	The large number of parameters can be reduced straightforwardly: 
	\begin{itemize}
		\item $a_{00} = F_M(0,0)$, low energy constraint.
		\item $a_{01} = (2/3)b_{11} F_M^{\infty}$, symmetric limit.
		\item $a_{10} = b_{20}(1+\delta_{BL})F_M^{\infty}$, fully asymmetric limit.
	\end{itemize}
	It has been seen that the pion TFF, $x F_\pi(x,0)$, marginally exceeds its asymptotic limit in the domain $x \textgreater 20$ GeV$^2$~\cite{Raya:2015gva}  (subsequently recovering it as $x$ continues to grow). Thus, we have included a parameter $\delta_{BL}$ to improve the quality of the fit for our given set of numerical data. This is by no means an implication that the Brodsky-Lepage limit of Eq.~\eqref{eq:BLlimit1} is violated; it is rather a numerical artifact to obtain a better interpolation. 
	
	The parametrization of Eq.~\eqref{eq:CAsdef} cannot be recast in any way so that the $Q_1 \cdot Q_2$ dependence in the $a_\mu^{\textrm{HLbL}}$ integrals disappears, but alternative methods can be implemented~\cite{Roig:2014uja,Masjuan:2017tvw}. Unlike LMD+$V_{N-1}$ parametrization, it also has a well defined limit when one of the variables is finite (but not zero) and the other tends to infinity. Since the large $Q^2$ domain is well under control, it enhances its reliability even far beyond the domain that contributes the most to $a_\mu^{\textrm{HLbL}}$, and that of the available data set.
	
	\subsection{LMD: $\eta_c$ and $\eta_b$}
	As explained in Ref.~\cite{Raya:2016yuj}, the $\eta_{c,b}$ TFFs lie below their corresponding asymptotic limits even at very large values of momentum transfer~\footnote{In a less noticeable way, this also happens for the $\eta'$. Thus, we found convenient to redefine $F_{\eta'}^\infty \to x_0 F(x_0,0)$, where $x_0=70$ GeV$^2$. We note that the symmetric form factor is not affected by this. Particularly, it is recovered exactly, to our numerical precision, for $x=y=10$ GeV$^2$.}. Imposing any asymptotic constraint is useless and potentially harmful for the accuracy of the fit. Moreover, those form factors are harder due to the larger masses of the $\eta_{c,b}$ mesons. In fact, the curvature of $\eta_b$ TFF is only very pronounced above a couple of hundred GeV$^2$~\cite{Chen:2016bpj}. 
	
	Therefore, a simple LMD-like form can be employed:
	\begin{equation}
	F_{\eta_c(\eta_b)}(x,y) = \frac{c_{00}+c_{10}(x+y)}{(x+M_{V_0}^2)(y+M_{V_0}^2)}\;,
	\end{equation}
	where we find $M_{V_0} := 3.097$ GeV $=m_{J/\psi}$, $c_{00}=6.5613$ GeV$^3$ and $c_{10}=0.0611$ GeV for $\eta_c$;
	$M_{V_0} := 9.460$ GeV$=m_\Upsilon$, $c_{00}=30.8424$ GeV$^3$ and $c_{10}=0.0426$ for $\eta_b$. Here the flaws of the large-$Q^2$ behavior of the LMD parametrizations are irrelevant: those appear far beyond the domain of integration. Notably, the LMD representation of the $\eta_b$ TFF, for $x,y\textless 20$ GeV$^2$, reproduces the numerical result within $1\%$ error. In the next section we present our numerical results for $a_\mu$.

	\section{Results}\label{Results}
\begin{figure}[t]
		\centering
		\begin{tabular}{c}
			\includegraphics[width=0.48\textwidth]{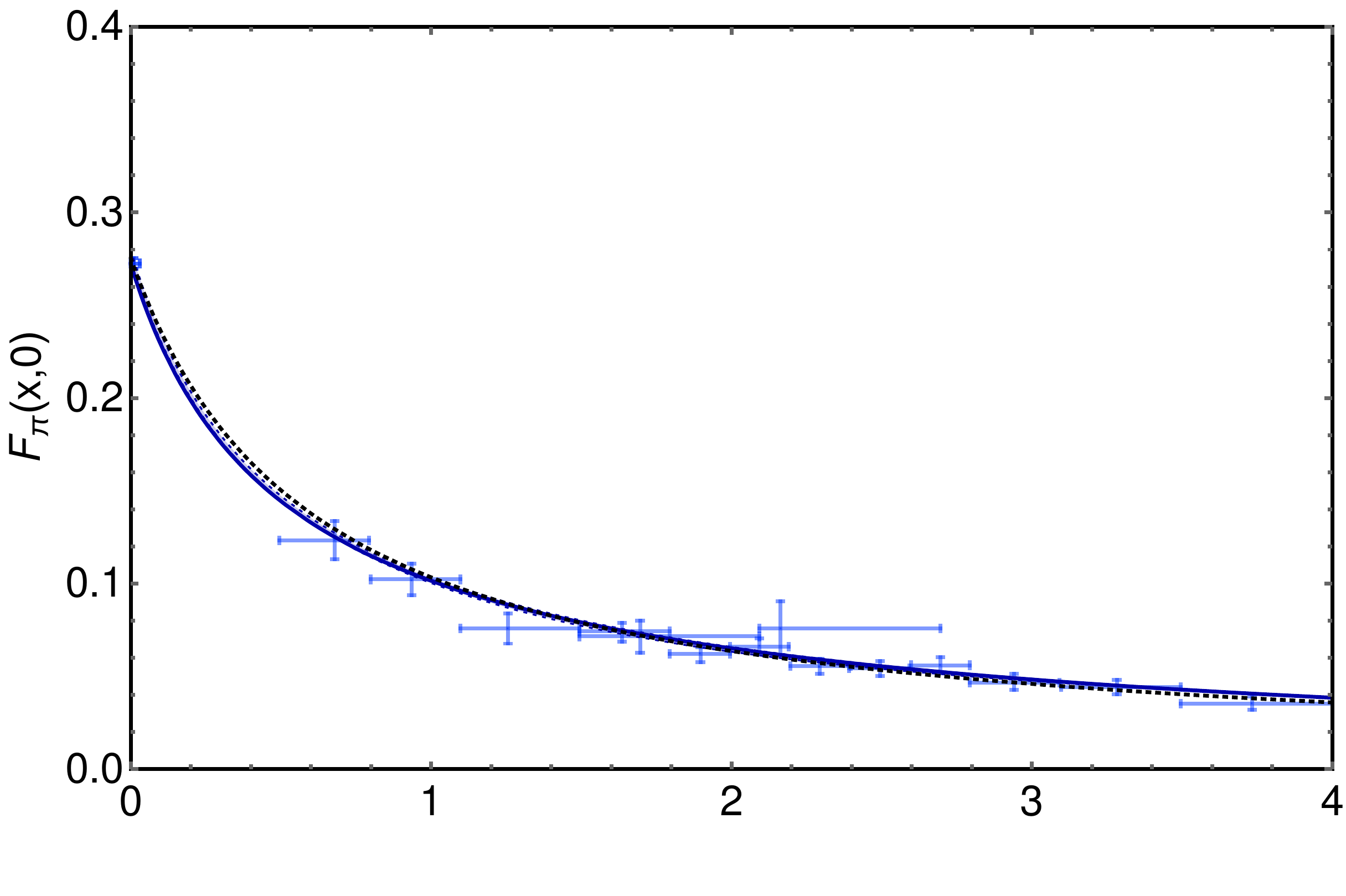}\\
			\includegraphics[width=0.48\textwidth]{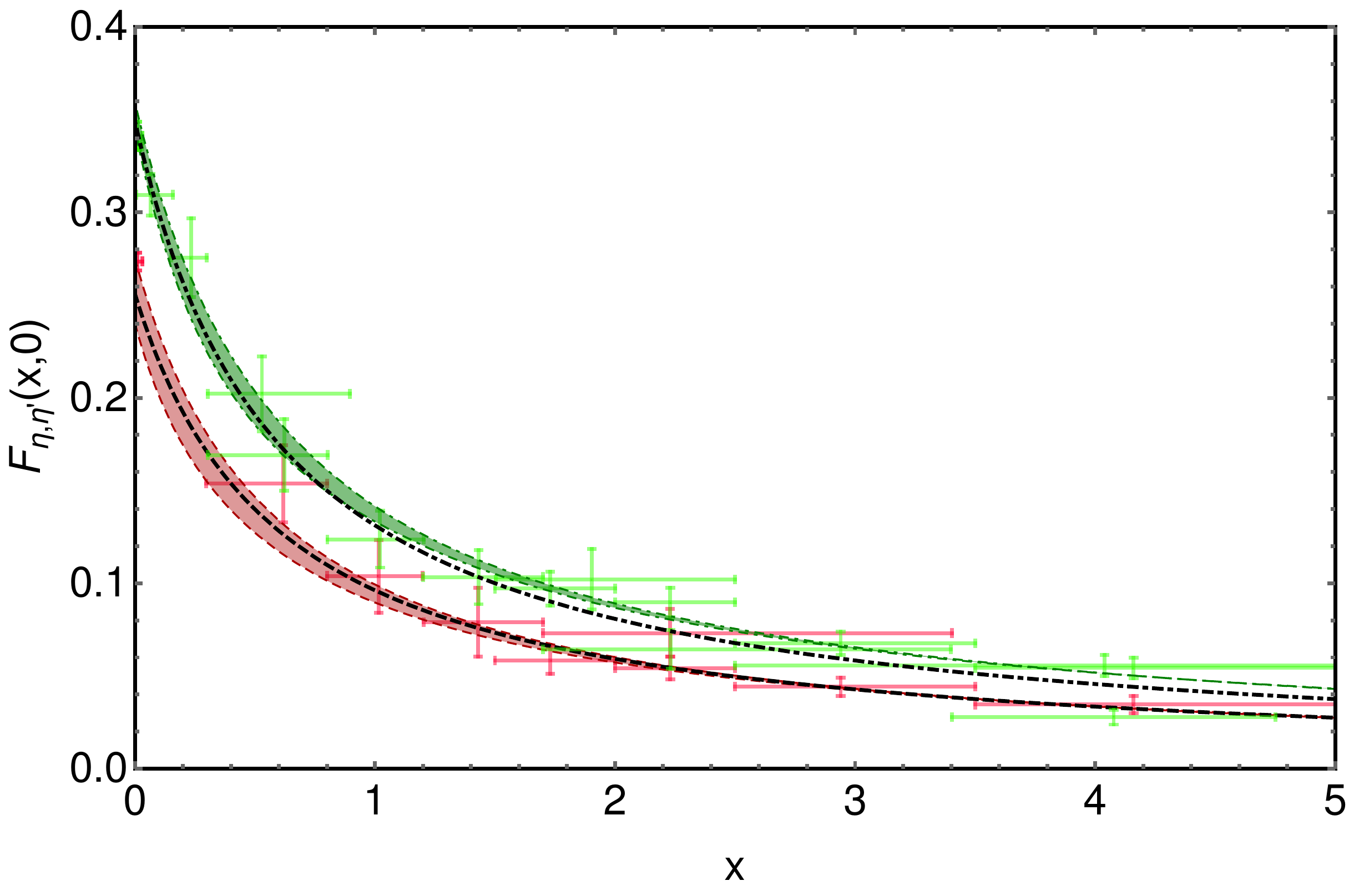}
		\end{tabular}
		\caption{[\textbf{Upper panel}] $\gamma \gamma^* \to \pi^0$ (solid curve). [\textbf{Lower panel}] $\gamma \gamma^* \to \eta,\eta'$. The band delimited by dashed and dot-dashed lines corresponds to $\eta$ and $\eta'$ TFFs, respectively, with the associate uncertainties. The dotted, dashed and dot-dashed curves are their corresponding VMD representations ($m_V=0.775$ GeV). Our choice of low-energy experimental data includes: CELLO~\cite{Behrend:1990sr} and CLEO~\cite{Gronberg:1997fj} collaborations (we have also included L3 data~\cite{Acciarri:1997yx} for the $\eta'$). Additionally, we display the most recent $x = 0$  values from PDG~\cite{Tanabashi:2018oca,Danilkin:2019mhd}. 
			The mass units are in GeV.}
		\label{fig:pionTFF}
		\centering
	\end{figure}

\begin{figure}[t]
		\centering
		\includegraphics[width=0.48\textwidth]{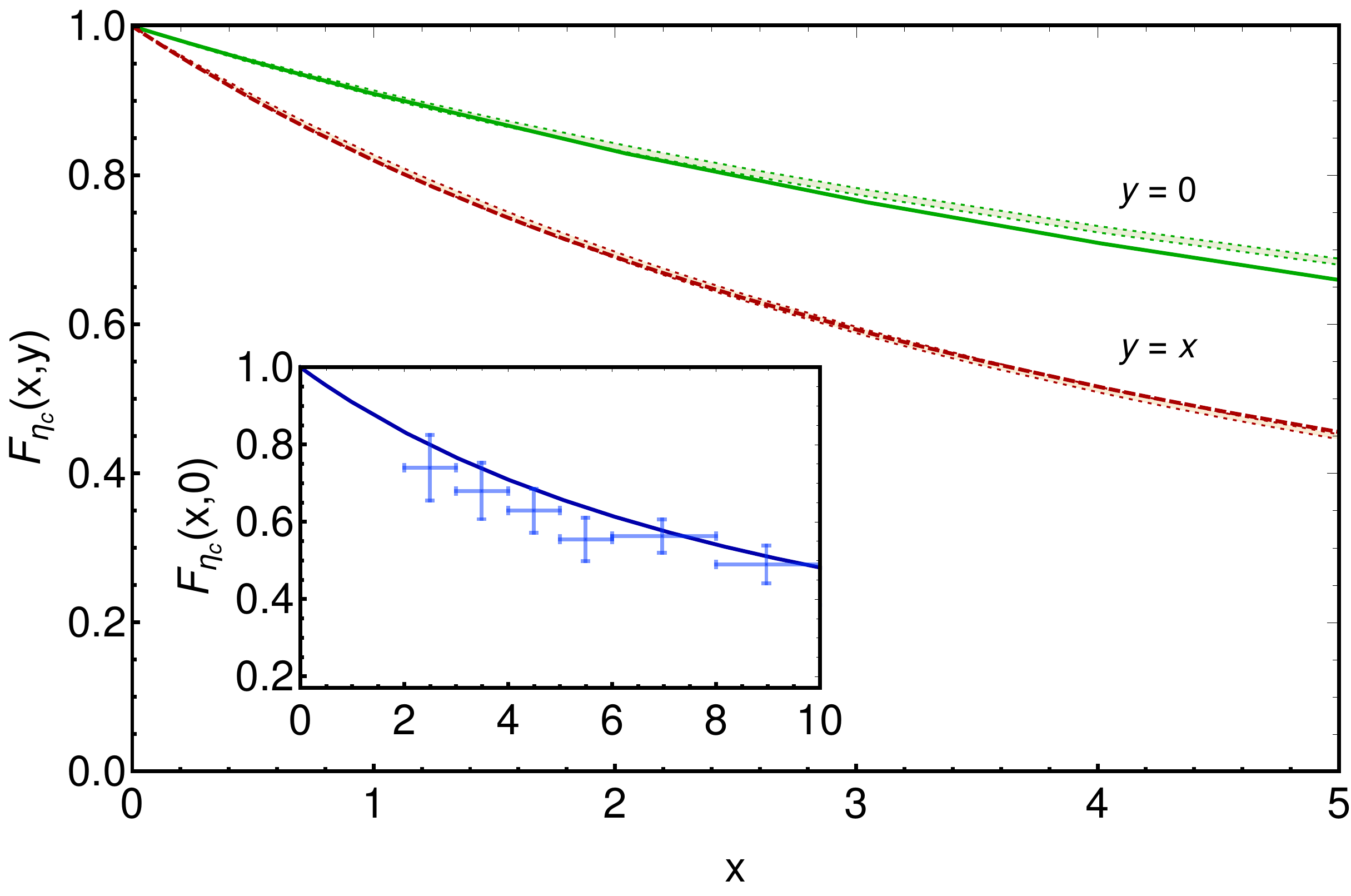}
		\caption{$\gamma^* \gamma^* \to \eta_c$ TFFs. 
			The (green) solid line corresponds to the direct numerical calculation of the SoF TFF, while the (red) dashed line is the analogous to the EoS case. The narrow bands are the corresponding results from our LMD representation.
			In the embedded plot, we compare our SDE prediction of $\gamma^* \gamma \to \eta_c$ with the available experimental data from BABAR~\cite{Lees:2010de}. The form factors have been normalized to unity. The mass units are in GeV. }
		\label{fig:heavy}
		\centering
	\end{figure}

\begin{figure}[t]
		\centering
		\includegraphics[width=0.48\textwidth]{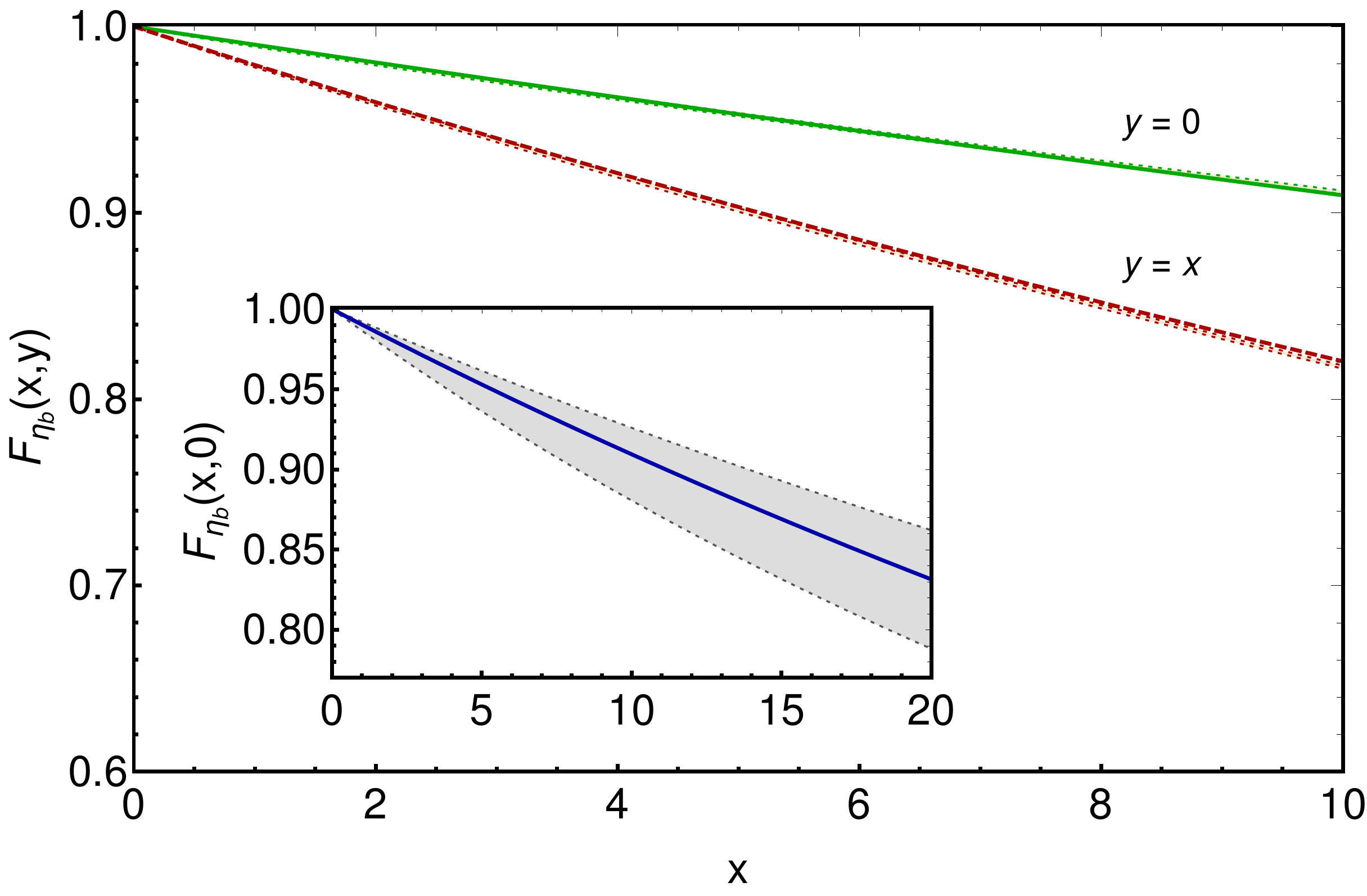}
		\caption{$\gamma^* \gamma^* \to \eta_b$ TFFs. 
			The (green) solid line corresponds to the direct numerical calculation of the SoF TFF, while the (red) dashed line is the analogous to the EoS case. The narrow bands are the corresponding results from our LMD representation.
			In the embedded plot, we compare our SDE prediction of $\gamma \gamma^* \to \eta_b$ with the nrQCD calculation from~\cite{Feng:2015uha} (gray band). The form factors have been normalized to unity. The mass units are in GeV. }
		\label{fig:heavy2}
		\centering
	\end{figure}

\begin{figure}[t]
		\centering
		\includegraphics[width=0.5\textwidth]{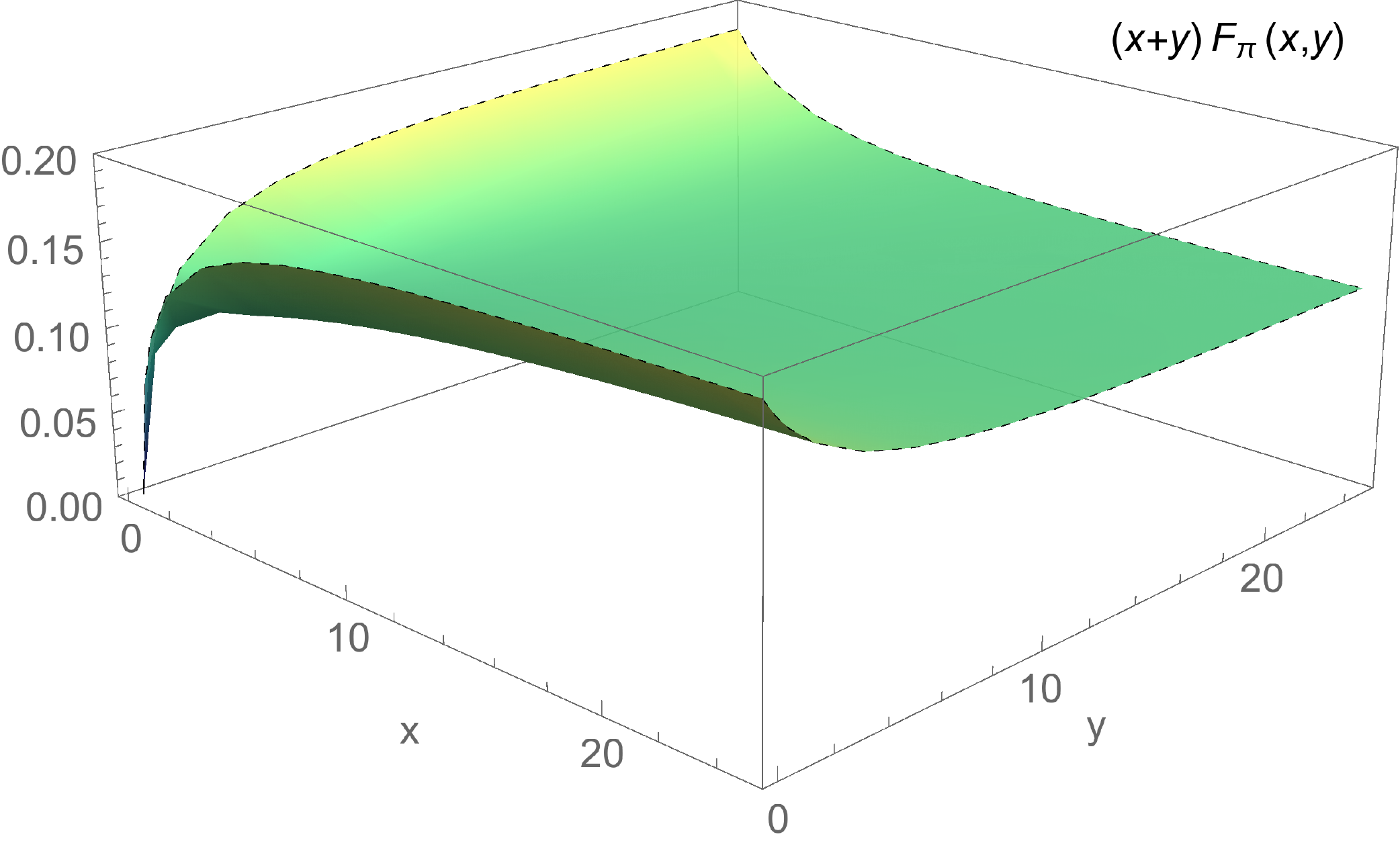}
		\caption{$\gamma^*\gamma^* \to \pi^0$ TFF. The mass units are in GeV.}
		\label{fig:pion3D}
		\centering
	\end{figure}

	We display the $\gamma \gamma^* \to \pi^0,\;\eta,\;\eta'$ TFFs in Fig.~\ref{fig:pionTFF} and their respective comparisons with available low-energy experimental data~\cite{Behrend:1990sr,Gronberg:1997fj,Acciarri:1997yx,Tanabashi:2018oca,Danilkin:2019mhd}; a keen agreement is exhibited. The $\eta_c$ result is plotted in Fig.~\ref{fig:heavy} and the analogous for $\eta_b$ in Fig.~\ref{fig:heavy2}. It is seen that the $\eta_c$ prediction matches the experimental data~\cite{Lees:2010de} and the corresponding for $\eta_b$ is in accordance with the non-relativistic QCD (nrQCD) approach~\cite{Feng:2015uha}. In the domain of interest, the LMD representations of $\eta_c$ and $\eta_b$ TFFs accurately reproduce the numerical SDE calculations.
	The corresponding results for all the pseudoscalars till much higher values of the probing photon momentum can be consulted in Refs.~\cite{Raya:2015gva,Raya:2016yuj,Ding:2018xwy}. Notably, the DoS extension we present in this work ensures $F_\pi(x,0)$ and $F_\pi(x,x)$ converge to their well-defined asymptotic limits, as can be observed in  Fig.~\ref{fig:pion3D}. The CAs 
	faithfully accommodate this numerical behavior. Additionally, the charge radius is reproduced to 1.5 \% accuracy.
	
	As we have discussed, all the pieces in our SDE-BSE treatment, in particular the QPV Ansatz, ensure an accurate description of the $\pi^0$, $\eta_c$ and $\eta_b$ mesons; but the case of $\eta-\eta'$ is not completely satisfactory. This occurs mostly due to the presence of the non-Abelian anomaly which, in principle, could introduce infrared corrections to the impulse approximation~\cite{Ding:2018xwy}.
	To account for the influence of the model for the QPV and other assumptions, firstly we vary the strength of the transverse terms in the QPV such that: 1) we reproduce (as much as possible) the empirical values of $F_{\eta,\eta'}(0,0)$ and 2) a rather large uncertainty is included in the more sensitive domain, around $Q^2 \sim 0.4$ GeV$^2$. From the computed decay constants, one gets a value of $F_\eta(0,0)$ which is about $10\%$ smaller than the empirical one, thus producing a broader band for the $\eta$ meson. In the case of $\pi^0$ and $\eta_c$, without the presence of the non-Abelian anomaly, the goal of this minimal variation is to produce an error band in the vicinity of $Q^2=0$, such that the uncertainty associated with $F_{\pi,\eta_c}(0,0)$ is comparable in size to that reported in PDG~\cite{Tanabashi:2018oca}. The $\eta_b$ TFF has not been measured yet. To be on the safe side, we include error bars on the charge radius by resorting to the nrQCD result. This produces a 5$\%$ error around $F_{\eta_b}(0,0)$.
	Any additional but reasonable change in the Bethe-Salpeter kernel parameters has a sufficiently small impact~\cite{Chen:2018rwz,Eichmann:2019tjk} and we find it to be contained within those bands. Furthermore, small variations of the meson masses also have negligible effects on the TFFs (in fact, one can take $m_\pi = 0$ with impunity), but they exhibit moderate to large impact on $a_\mu$. Thus we allow ourselves to vary $m_\pi \sim 0.135\; - \;0.140$ GeV, $m_\eta \sim 0.548 \;-\;0.560 $ GeV and $m_{\eta'} \sim 0.956\;-\;0.960$ GeV.

	Putting all together, we obtain:
	\begin{eqnarray}
	\nonumber
	a_\mu^{\pi^0-\textrm{pole}}&=&\left(6.14\pm0.21 \right)\times 10^{-10}\;,\\
	\nonumber
	a_{\mu}^{\eta-\textrm{pole}} &=& (1.47 \pm 0.19) \times 10^{-10}\;, \\
	\nonumber
	a_{\mu}^{\eta'-\textrm{pole}} &=& (1.36 \pm 0.08) \times 10^{-10}\;, \\
	\nonumber
	a_{\mu}^{\eta_c-\textrm{pole}} &=& (0.09 \pm 0.01) \times 10^{-10}\;, \\
	a_{\mu}^{\eta_b-\textrm{pole}} &=& (0.26 \pm 0.01) \times 10^{-13}\;.
	\end{eqnarray}
	Our SDE prediction of $a_\mu^{\pi^0-\textrm{pole}}$ is compatible with  other reported values~\cite{Knecht:2001qf,Masjuan:2017tvw, Guevara:2018rhj,Roig:2014uja,Hoferichter:2018dmo, Hoferichter:2018kwz}. For example, $a_\mu^{\pi^0-\textrm{pole}}= (5.81\pm0.09\pm0.09{}^{+0.5}_{-0}) \cdot10^{-10}$ according to Ref.~\cite{Guevara:2018rhj} (resonance chiral lagrangians), while Hoferichter \textit{et al.} obtain $a_\mu^{\pi^0-\textrm{pole}}= (6.26^{+0.30}_{-0.25}) \cdot10^{-10}$ (dispersive evaluation). Ref.~\cite{Masjuan:2017tvw} reports $a_\mu^{\pi^0-\textrm{pole}}= (6.36\pm0.26) \cdot10^{-10}$ (CAs) and a recent SDE evaluation~\cite{Eichmann:2019tjk} obtains $a_\mu^{\pi^0-\textrm{pole}}= (6.26\pm0.13) \cdot10^{-10}$.
	
	From the $\eta-\eta'$ pole contributions, our full-SDE results yield $a_\mu^{\eta+\eta'-\textrm{pole}}=(2.83 \pm 27)\cdot10^{-10}$.
	
	Assuming a two-angle mixing scheme in the flavour basis~\cite{Feldmann:1999uf}, and a chiral approach~\cite{Roig:2014uja}:
	\begin{eqnarray*}
		F_s(x,y) = F_{l=u,d}(x,y) = F_\pi(x,y)\;,
	\end{eqnarray*}
	one can write the $\eta-\eta'$ TFFs in terms of $F_\pi(x,y)$. This simplification yields $a_\mu^{\eta+\eta'-\textrm{pole}}=(2.86\pm0.42)\cdot10^{-10}$, which is consistent with our result albeit with a larger error. The SDE result from Ref.~\cite{Eichmann:2019tjk} follows a symmetry-preserving RL approach to compute $F_l(x,y)$ and $F_s(x,y)$, such that the physical states $\eta-\eta'$ are obtained from there after assuming a two-angle mixing scheme. It is shown that whether one takes the chiral approach or not, the sum of the $\eta-\eta'$ contributions to $a_\mu$ remains the same (although the individual contributions are different). This is due to cancellations that occur because of the structure of the mixing matrix. However, in our SDE-BSE treatment, the mixing between the $l$ and $s$ flavours is produced directly due to the presence of the non-Abelian anomaly kernel in the BSE, Eq.~\eqref{eq:anomaly}. It is the anomaly kernel that produces the mixing; no particular mixing scheme is assumed. For our data sets, limited to the range $x,y\leq 10$ GeV$^2$, we show the CAs parameters in Table~\ref{Tab:CAsparams}.
	
	\begin{table}[htbp]
		\caption{\label{Tab:CAsparams} CAs parameters, from Eq.~\eqref{eq:CAsdef}, of $F_M(0,0)$ (which has units of GeV$^{-1}$ and the parameters have units accordingly). For the pion, $\delta_{BL}=0.0437$ and $\delta_{BL}=0$ in the other cases. }
		\begin{center}
			\begin{tabular}{@{\extracolsep{0.0 cm}}crrrr}
				\hline \hline
				Meson & $b_{01}$ & $b_{10}$ & $b_{11}$ & $b_{20}$  \\
				\hline
				$\pi^0$ & $6.1301$ & $2.7784$ & $0.2147$ & $1.1301$  \\
				$\eta$ & $14.5769$ & $4.1981$ & $0.4323$ & $3.7460$ \\
				$\eta'$ & $5.3256$ & $2.6822$ & $0.0245$ & $1.0933$ \\
				\hline \hline
			\end{tabular}
		\end{center}
	\end{table}
	
	Regarding the heavy mesons, although the value of $\eta_b$ is 3 orders of magnitude smaller, the $\eta_c$ is commensurate with the current experimental and theoretical error bars. Moreover, our obtained value is fully compatible with that reported in Ref.~\cite{Colangelo:2019uex}, $a_{\mu}^{\eta_c-\textrm{pole}} = (0.08) \cdot 10^{-10}$. Thus, this contribution might not be omitted when the theoretical calculations reach a higher level of precision. Also, it could serve as an estimate for potential non-perturbative corrections to the charm loop~\footnote{We thank P. Masjuan, P. Sanchez-Puertas and M. Hoferichter for discussions on the topic and confirming to us  that they reached compatible values for the $\eta_c$ contribution.}.
	
	\section{Conclusions}\label{Concl}
	It is highly timely to revisit the computation of $a_\mu^{\textrm{HLbL}}$ on the eve of the FNAL (and hopefully J-PARC) improved measurements. We calculate the dominant piece of this observable (coming mainly from the $\pi^0$ pole and secondarily from the $\eta$ and $\eta'$ poles). For the first time, the sub-leading contributions of $\eta_c$ and $\eta_b$ poles were obtained. As a result of our analysis, we find:
	\begin{eqnarray}
	\nonumber
	a_\mu^{\pi^0-\textrm{pole}}&=&\left(6.14\pm0.21 \right)\times 10^{-10}\;,\\
	\nonumber
	a_{\mu}^{\textrm{\emph{light}}-\textrm{pole}} &=& (8.97 \pm 0.48) \times 10^{-10}\;, \\
	\nonumber
	a_{\mu}^{\textrm{\emph{all}}-\textrm{pole}} &=& (9.06 \pm 0.49) \times 10^{-10}\;. 
	\end{eqnarray}
	Our findings for the light pseudoscalars are compatible with previous determinations and have a comparable uncertainty. While the $\eta_b$ result is negligible, the magnitude of $a_{\mu}^{\eta_c-\textrm{pole}}$ (confirmed in Ref.~\cite{Colangelo:2019uex}) is sizable as compared with the contemporary error bars; thus, it could promote more theoretical calculations on the topic.

	Earlier and recent SDE works~\cite{Goecke:2010if,Eichmann:2019tjk,Eichmann:2019bqf} have shown this continuum approach as a promising tool  in understanding the QCD contributions to $a_\mu$.  This is clearly supported by the consistency with our predictions and those from~\cite{Eichmann:2019tjk}. Moreover, the present work heavily relies on our earlier studies,~\cite{Raya:2015gva,Raya:2016yuj,Ding:2018xwy}, where we
	compute the pseudoscalar transition form-factors: $\gamma \gamma^* \to \{\pi^0, \eta, \eta', \eta_c, \eta_b\} $, all lowest-lying neutral pseudoscalars.  
	Such calculations are based upon a systematic and unified treatment of QCD's SDEs. Several efforts  have followed this approach to compute a plethora of hadron properties, with the resulting predictions invariably being in agreement with or confirmed by experimental data and lattice QCD simulations (see Refs.~\cite{Aguilar:2019teb,Horn:2016rip} for recent reviews). 
	
	Our previous research~\cite{Raya:2015gva,Raya:2016yuj,Ding:2018xwy} and the resulting current work not only explain the existing data accurately but are also quantitatively predictive for the ones to be measured in modern facilities. Thus, we believe this work is useful in the collective effort to reduce the error of the SM prediction of $a_\mu$ so as to maximally benefit from the forthcoming improved measurements and \emph{hopefully} find indirect evidence for new physics in the future. 
	
	\section*{Acknowledgements}
	K. Raya wants to acknowledge L. Chang, M. Ding and C. D. Roberts for their scientific advice.
	P. Roig thanks P. Masjuan and P. S\'anchez-Puertas for useful conversations on this topic. This research was also partly
	supported by Coordinaci\'on de la Investigaci\'on Cient\'ifica
	(CIC) of the University of Michoacan, CONACyT-Mexico and SEP-Cinvestav, through Grant nos. 4.10, CB2014-22117, 
	CB-250628, and 142 (2018), respectively.
	
	\appendix
	
	\section{Quark propagator and BS amplitudes}\label{appendix}
	It is convenient to express the quark propagator in terms of complex conjugate poles (ccp). Omitting flavor indices, it can be expressed as
	\begin{eqnarray}
	\nonumber
	S(p)&=& -i\gamma \cdot p\;\sigma_v(p^2)+\sigma_s(p^2)\\
	&=&\sum^{j_m}_{j=1}\left[\frac{z_j}{i\gamma{\cdot}p+m_j}+\frac{z^{*}_j}{i\gamma{\cdot}p+m^{*}_j}\right],\label{quarkPTIR}
	\end{eqnarray}
	where $z_j,m_j$ are obtained from a best fit to the numerical solutions, ensuring \textrm{Im}$(m_j)\neq 0\forall j$, a feature consistent with confinement~\cite{Chang:2013pq}. We find that $j_m=2$ is adequate to provide an accurate interpolation.
	
	The BS amplitude of a neutral pseudoscalar is written in terms of four covariants, namely: 
	\begin{eqnarray}\nonumber
	\Gamma_M(k;P) &=& \gamma_5\{ \mathcal{F}_1+ \gamma \cdot P\; \mathcal{F}_2+(k\cdot P)\gamma \cdot P \;\mathcal{F}_3\\
	&+& i[\gamma \cdot k,\gamma \cdot P] \;\mathcal{F}_4\}
	\end{eqnarray}	
	
	Each scalar function, $\mathcal{F}_k=\mathcal{F}_(k;P)$, is split in two parts and can be parametrized in terms of PTIRs in the following way:
	\begin{subequations}\label{BSAPTIR}
		\begin{eqnarray}
		\mathcal{F}(k;P)&&=\mathcal{F}^i(k;P)+\mathcal{F}^u(k;P)\,,\\
		\mathcal{F}^i(k,P)&&=c^i_{\mathcal{F}}\int^1_{-1}dz\rho_{\nu^i_{\mathcal{F}}}(z)\big[a_{\mathcal{F}}\widehat{\Delta}^4_{\Lambda^i_{\mathcal{F}}}(k^2_z)\notag\,,\\
		&&               +a^-_{\mathcal{F}}\widehat{\Delta}^5_{\Lambda^i_{\mathcal{F}}}(k^2_z)\big]\,,\\
		\mathcal{F}^u(k;P)&&=c^u_{\mathcal{F}}\int^1_{-1}dz\rho_{\nu^u_{\mathcal{F}}}(z)\widehat{\Delta}^{{l^u_{\mathcal{F}}}}_{\Lambda^u_{\mathcal{F}}}(k^2_z)\,.
		\end{eqnarray}
	\end{subequations}
	with $\widehat{\Delta}_{\Lambda}(s)=\Lambda^2\Delta_\Lambda(s)$, $k^2_z=k^2+zk{\cdot}P$, $a^-_\mathcal{F}=1-a_\mathcal{F}$. The indices `$i$' and `$u$' denote the connection with the infrared and ultraviolet behaviors of the BS amplitude and, the spectral density:
	\begin{equation}
	\rho_\nu(z)=\frac{\Gamma\left[\frac{3}{2}+\nu\right]}{\sqrt{\pi}\Gamma\left[1+\nu\right]}(1-z^2)^\nu
	\end{equation}
	The interpolating parameters are obtained through fitting to the Chebyshev moments:
	\begin{align}
	\mathcal{F}_n(k^2)=\frac{2}{\pi}\int^1_{-1}dx\sqrt{1-x^2}\mathcal{F}(k;P)U_n(x),
	\end{align}
	with $n=0,2$, where $U_n$ is an order-$n$ Chebyshev polynomial of the second kind. $\mathcal{F}_4(k;P)$ is small and has no impact, hence it is omitted in all cases. For similar reasons,  $\mathcal{F}_3(k;P)$ might be omitted for $\eta'$ and $\eta_b$ as well.
	
	The forms given in Eqs.\eqref{quarkPTIR}-\eqref{BSAPTIR} have been proven undoubtedly useful; their specific interpolating values are presented in Refs.~\cite{Raya:2015gva,Raya:2016yuj,Ding:2018xwy}.

	\bibliographystyle{unsrt}
	\bibliography{main}

\end{document}